# A Low-Frequency, Autoresonant Wireless Power Transfer Link for Bidirectional Bionic Interfaces

Giulio Maria Bianco
*Prosthetic Center of Budrio*
*Italian Workers' Compensation Authority*
Vigorso di Budrio, Budrio, Italy
ORCID: 0000-0002-3216-5884

Alberto Dellacasa Bellingegni
*Prosthetic Center of Budrio*
*Italian Workers' Compensation Authority*
Vigorso di Budrio, Budrio, Italy
ORCID: 0000-0002-9569-9680

Federico Mereu
*Prosthetic Center of Budrio*
*Italian Workers' Compensation Authority*
Vigorso di Budrio, Budrio, Italy
ORCID: 0000-0002-5835-2354

Daniel Gelmini
*Rehab Technologies Lab*
*Istituto Italiano di Tecnologia*
Genoa, Italy
daniel.gelmini@iit.it

Michele Canepa
*Rehab Technologies Lab*
*Istituto Italiano di Tecnologia*
Genoa, Italy
michele.canepa@iit.it

Matteo Laffranchi
*Rehab Technologies Lab*
*Istituto Italiano di Tecnologia*
Genoa, Italy
matteo.laffranchi@iit.it

Emanuele Gruppioni
*Prosthetic Center of Budrio*
*Italian Workers' Compensation Authority*
Vigorso di Budrio, Budrio, Italy
ORCID: 0000-0003-0732-8378

***Abstract*—To provide multimode sensory feedback and motion control, bidirectional bionic interfaces for advanced prosthetic systems require continuous and secure energy delivery to implantable electronics and integration in the sensing WBAN (Wireless Body Area Network) of the patient. However, powering such interfaces is still an open issue. Wireless Power Transfer (WPT) avoids implanted batteries and transcutaneous connections, but its design is constrained by stringent requirements on electromagnetic safety, implant size, voltage compliance, and coexistence with sensitive bio-signal acquisition and stimulation circuitry. This paper presents the design and testing of a low-frequency (127 kHz) inductive WPT link for an implantable bidirectional bionic interface. The system includes an autoresonant driving control to maintain operation at resonance under varying coupling and load conditions of the cyber-physical prosthesis. Starting from the requirements of the bionic interface, the wireless body-area sensing system is designed by selecting the working frequency, drawing the electrical schemes, and checking its safety and regulatory compliance. Preliminary WPT prototypes can provide up to ~140 mA and ~20 V, achieving a maximum power transfer efficiency higher than 40% and satisfying the project requirements up to a 2 cm implantation depth.***

## I. INTRODUCTION

*Bionic interfaces* are systems designed to establish a functional and reliable communication pathway between the human body and artificial devices similarly to the biological interfaces [1]. To enable the simultaneous exchange of control signals and sensory information, such human-machine interfaces must support bidirectional and multi-modal information transfer, viz., a two-way communication interacting with multiple signal types such as electromyographic (EMG) and electroneurographic (ENG) signals [2, 3]. In the field of prosthetics, bionic interfaces can enable the restoration of lost motor functions and sensory feedback, with the goal of achieving natural interactions between the user and the prosthetic device [1]. Consequently, they constitute complex cyber-physical systems encompassing biological, mechanical, electrical, and software components.

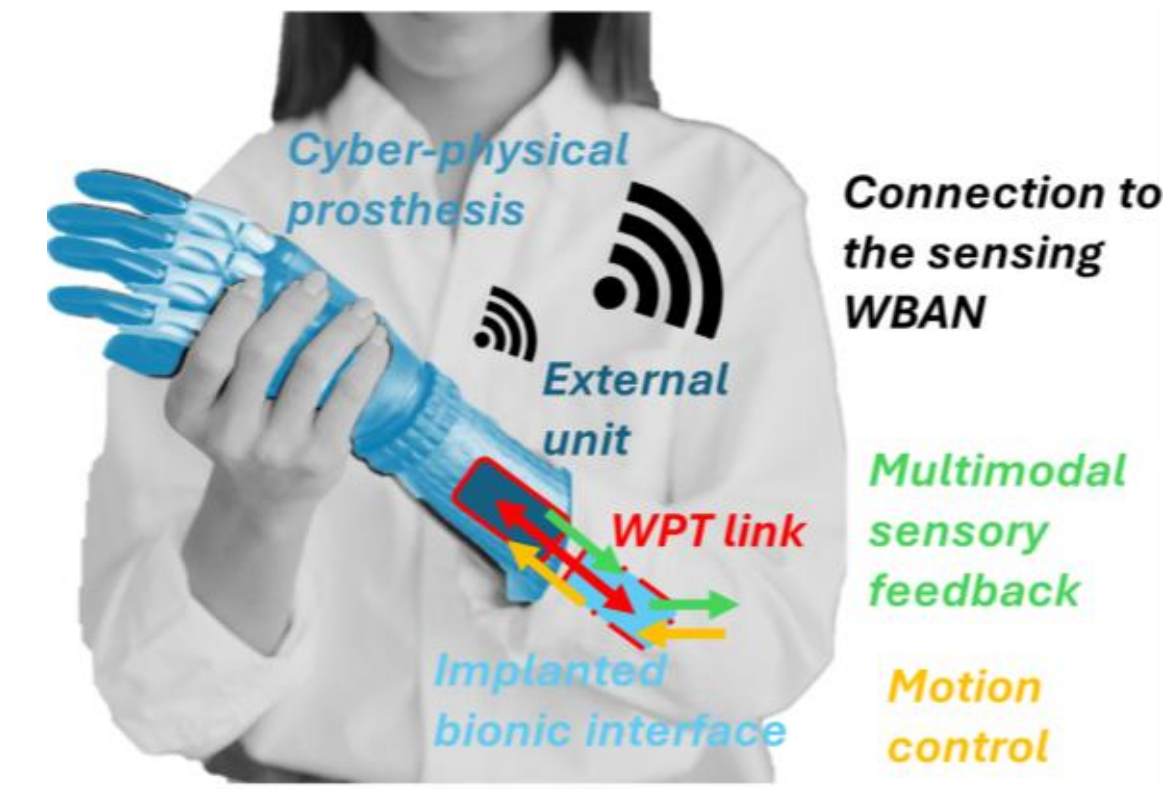

Fig. 1. Concept of a cyberprosthesis with bidirectional bionic interface and wireless connectivity powered through Wireless Power Transfer (WPT).

Modern prosthetic limbs rely on increasingly sophisticated actuators, sensors, and embedded electronics [4]. However, their overall performance and clinical acceptance are often limited by the quality of the human–machine interface [5]. Traditional approaches, such as surface electromyography (sEMG)

This work was supported by the Italian Workers' Compensation Authority (INAIL) under Grant PR23-PAS-P2-BioInterNect (Interfacce bioniche bidirezionali multimodali; Bi-directional multi-mode bionic interfaces).

electrodes and percutaneous connectors, suffer from well-known drawbacks including signal instability, susceptibility to noise, skin irritation, infection risk, and limited long-term reliability [6]. These limitations motivate the development of implantable bionic interfaces capable of providing stable feedback between the prosthesis and the biological tissue, including both sensory feedback and motion control.

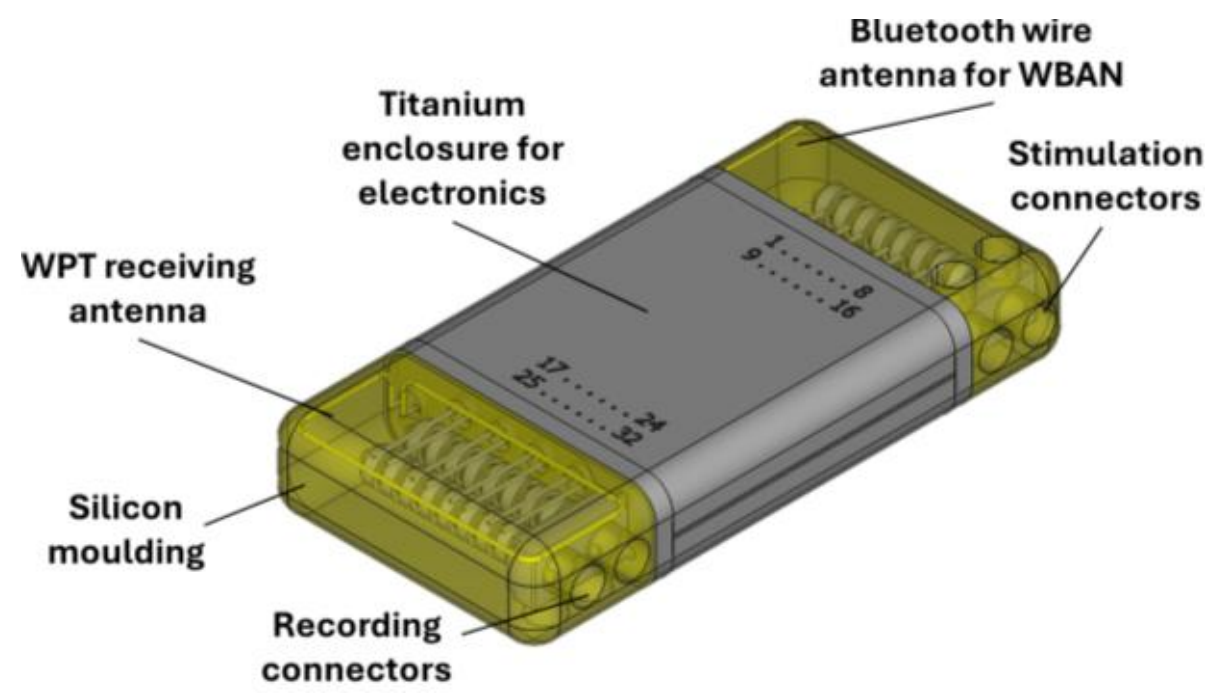

Fig. 2. Preliminary CAD of the implantable unit

However, powering such interfaces is still an issue to be addressed properly. Conventional solutions based on implanted batteries present significant limitations due to battery size and finite lifetime necessitating surgical replacement and/or transcutaneous wires. Moreover, bidirectional and multimodal interfaces typically require variable and sometimes peak power levels, which are difficult to sustain with miniaturized batteries alone. Thanks to recent development in biomedical wireless sensors [7] and systems [8], Wireless Power Transfer (WPT) can overcome these limitations by enabling continuous energy delivery from an external unit, eventually embedded in the prosthesis itself, to the implant, which interfaces with the hosting body and can connect to the WBAN (Wireless Body Area Network) of the patient (Fig. 1); the final goal is decoupling implant functionality from on-board energy storage [2].

Inductive, Low-Frequency (LF) antenna systems are a viable choice for testing such links thanks to the relatively weak interaction of the electromagnetic fields with the biological tissue at those frequencies [2, 9]. Naturally, the development of such a WPT system is complex and multifaceted [9, 10], because of *i*) electromagnetic safety and exposure limits, *ii*) limited power transfer efficiency, *iii*) implant size constraints, and *iv*) regulatory constraints. In this paper, the preliminary results of a LF WPT link for powering implanted bionic interfaces are presented, including the CAD of the goal device, the description of the antenna system, the safety and regulatory compliance, and the experimentally measured power transfer.

## II. Wireless System Design

In this Section, the antenna system is described including: *i*) the project requirements; *ii*) the operating frequency selection; the custom boards *iii*) transmitting and *iv*) receiving; and, lastly, *v*) the preliminary safety checks, to be completed

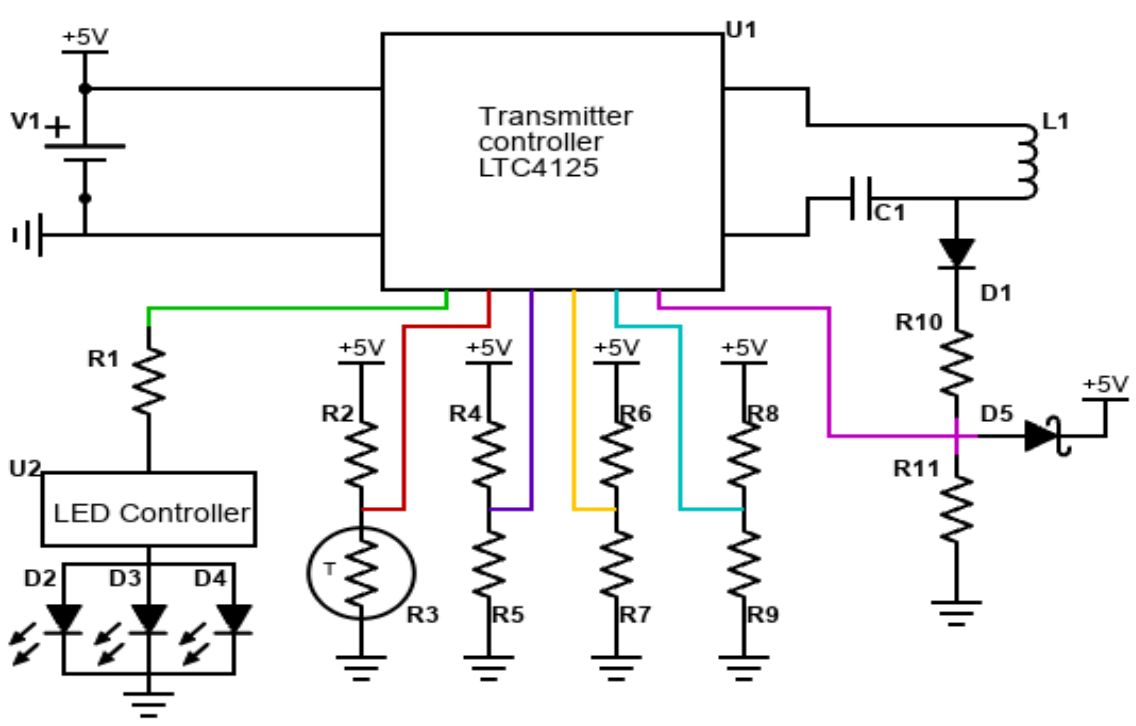

Fig. 3. Schematic of the Tx board with the details of the safety components.

with actual measurements from an approved certifier of medical devices.

### A. *Requirements of the Bionic Multimodal Interface*

The design of the WPT is driven by stringent electrical stimulation requirements. The possible implantation depth of the internal unit ranges from 5 mm to 20 mm. For invasive electrical stimulation, the implantable hardware must generate current-controlled stimulation pulses with amplitudes programmable from 0 mA up to 2 mA for typical operations, with a fine resolution of 0.5 µA; stronger sensory stimulation could require up to 5 mA of current. The stimulation pulse width must be configurable from 10 µs to 500 µs in steps of 10 µs, with a minimum interphase delay of 100 µs to support biphasic waveforms and charge-balanced stimulation. From a power supply perspective, the stimulation circuitry requires a voltage compliance of ±7 V to guarantee accurate current delivery across variable electrode–tissue impedances. The implantable system must support more than 8 independent stimulation channels operating concurrently, further increasing the aggregate power demand and the need for stable supply regulation. These requirements must be satisfied within strict form-factor constraints, with maximum dimensions limited to 45 mm × 45 mm × 10 mm. While the WPT powers the implantable unit, information transfer must be provided by another wireless link, which could use, for instance, the Bluetooth protocol (as investigated previously on this same topic [11]).

Fig. 2 shows a preliminary CAD of the implanted unit: electronics is embedded in a sealed titanium enclosure, with feedthroughs exposing the recording and stimulation signals towards inline connectors; the WPT coil is embedded in a silicon-moulded area on the left side of the enclosure. A Bluetooth wire antenna is embedded on the opposite side of the enclosure, to provide bidirectional communication to the patient's WBAN. In parallel, the bionic interface architecture includes a Transcutaneous Electrical Nerve Stimulation (TENS) [12] module intended for non-invasive stimulation of peripheral nerves; this module, to be driven with a booster of currents and voltages, is currently under development and outside the scope of this paper.

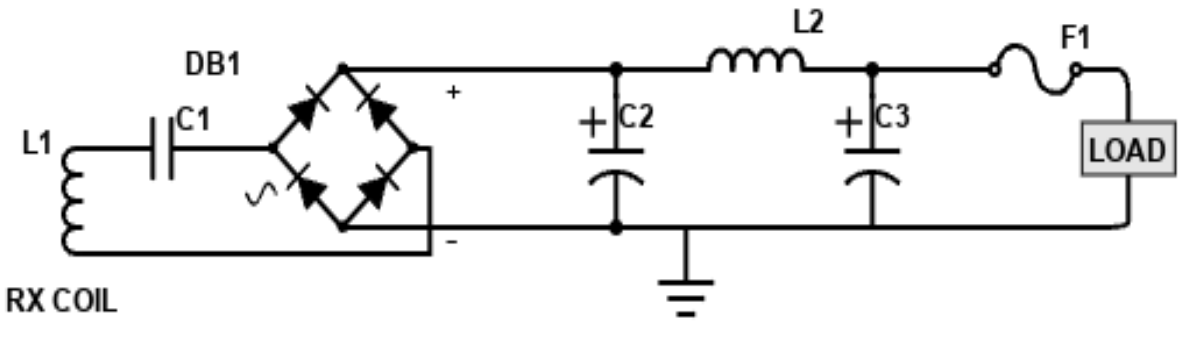

Fig. 4. Schematic of the Rx board.

Finally, the software and firmware architecture must support fully programmable stimulation waveforms, including at least square and sinusoidal modes, and must interface with the hardware through standard microcontroller-compatible input/output protocols. The software must enable independent management of waveform parameters for each stimulation channel, allow runtime modification of stimulation settings, and support concurrent stimulation with channel-specific parameters. Software-based security mechanisms must complement the hardware safeguards to ensure safe operation under all operating conditions.

### B. *Choice of the Working Frequency*

The first design choice in the antenna system is the working frequency. In the LF band, the frequency range 119-135 kHz in the European Union can be utilized for short-range systems like LF RFID tags [13]; these frequencies are reserved for professional terrestrial digital mobile radios with extended band, maritime mobile, and naval navigation systems based on impulses but can be employed for short-range, relatively low-power systems like the present one [14]. The safety limits on the Specific Absorption Rate (SAR) and induced E field inside the biological tissue are, furthermore, easy to meet at such a low frequency (see next Sec. II-E).

To compensate for possible frequency shifts due to deterioration of the components and/or differences in the patients' bodies, the centre frequency of 127 kHz is selected. The chosen implanted coil is the model 760308101303 (by Würth Elektronic; inductance: 47 µH) which has an external diameter of 26.3 mm and hence is compatible with the project specifications. The external coil is the model 760308100110 (by Würth Elektronic; inductance: 24 µH) having external diameter of 50.0 mm. The coils achieve resonance through capacitors connected in series with them according to the well-known equation

$$f_0 = \frac{1}{2\pi\sqrt{LC}} \qquad (1)$$

where $f_0$ is the resonant frequency, $L$ is the coil's inductance and $C$ is the capacitance connected in series to the coil.

### C. *Transmitting Circuit*

The transmitting circuit is based on the LTC4125 by Linear Technology [15] IC (Integrated Circuit). The LTC4125 IC is provided with an autoresonant feature, which maximizes the WPT similarly to the selftuning feature of UHF RFIDs [16, 17]. Such autoresonance works by constantly adjusting the excitation frequency to match the natural resonance of the LC tank. The transmitter drives the resonant network with a square-wave voltage rather than a sinusoidal source. When the quality factor Q of the resonant circuit is sufficiently high (Q>10), the frequency-selective behaviour of the LC tank strongly attenuates higher-order harmonics of the square wave, so that the resulting voltage and current waveforms across the inductor and capacitor are dominated by the fundamental component at the resonant frequency. At startup, the driver initially excites the LC tank with a fixed-frequency square wave to search for eventual inductive coupling. As current begins to circulate in the resonant network, the system detects the phase relationship between the switching node voltage and the inductor current. The driving frequency is then adjusted in real time to maintain a zero-phase difference between voltage and current, which corresponds to operation at resonance. This phase alignment is hence enforced on a cycle-by-cycle basis [15]. The autoresonance is a critical feature for providing always the maximum achievable power while the working conditions of the cyberprosthesis change, for instance, due to different loads or health conditions of the patient.

The simplified electrical scheme of the board is depicted in Fig. 3, wherein the different parts of the board are highlighted in colours. The green components return feedback on the transmitted power using LEDs. The next three circuits are safety checks: the red one monitors the temperature of the board; the purple one ensures that the maximum delivered power doesn't surpass a threshold; the yellow one checks if the resonant frequency is the expected one. The cyan circuit sets the minimum power employed for seeking the receiving board through autoresonance. Lastly, the magenta connection provides feedback (between the Tx coil, L1, and the resonance capacitor, C1) used by the IC for governing the autoresonance of the coil. The LTC4125 can meet all the requirements concerning the waveform and resolution; however, it must be controlled by custom software/firmware that are currently under development.

### D. *Receiving Circuit*

The receiving circuit of the bionic interface is responsible for managing the stimulation channels and the TENS. However, current prototypes shown in this work are still limited to basic functions and focused on the WPT development. Consequently, the Rx board used for testing the WPT is a simple PCB designed starting from previous research [18]. The scheme of the board is in Fig. 4. The resonant circuit is connected to a full-wave bridge rectifier, which is then connected to a LCL stabilizer and provides DC current to the load. For this experimentation, a motor is connected to the board, so that the continuous current drain used by the motor allows for testing the correct functioning of the autoresonant circuit.

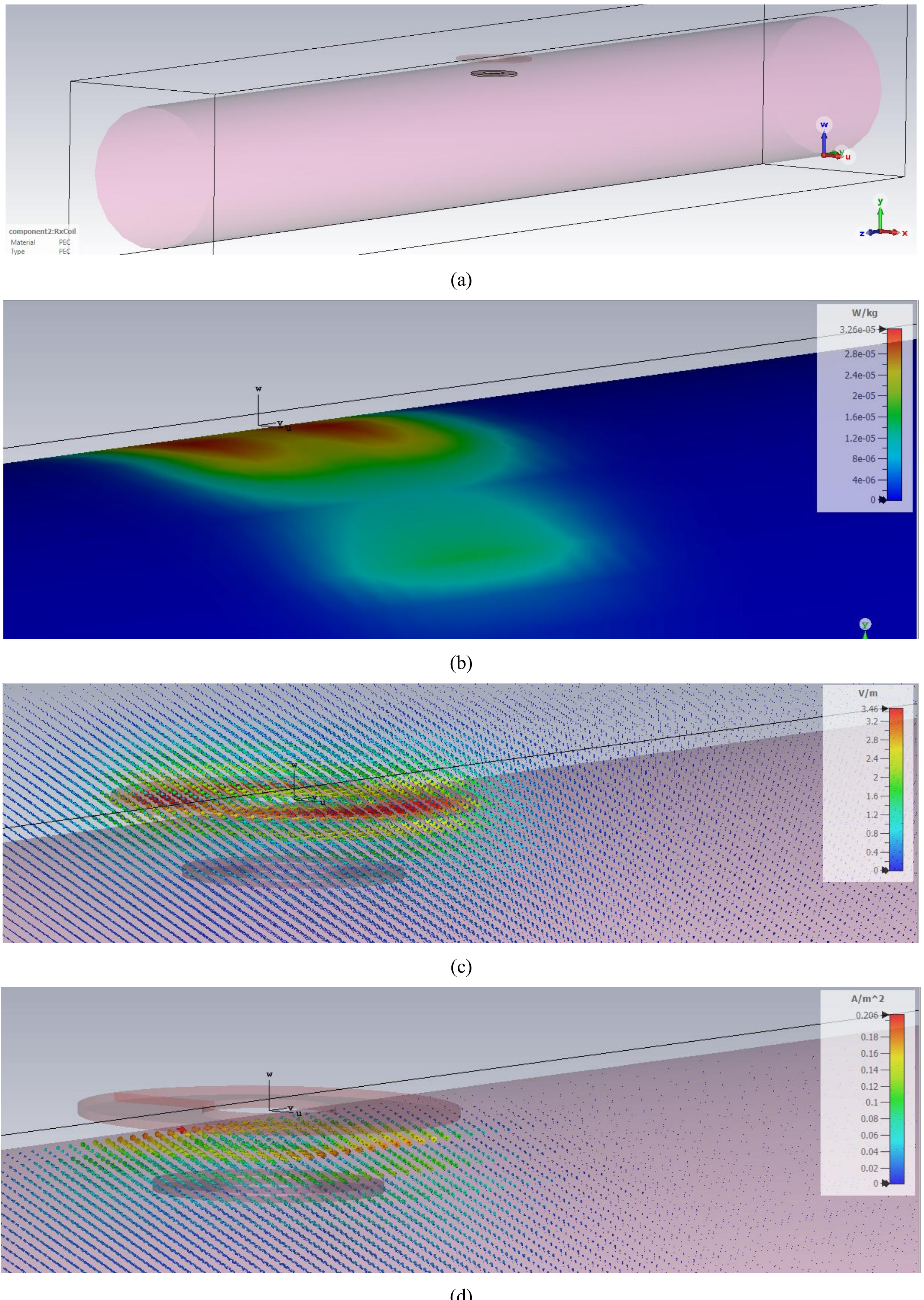

Fig. 5. Simulative check of the WPT safety. (a) CAD of the simulation. Computed values: (b) SAR, (c) electric field and (d) current density.

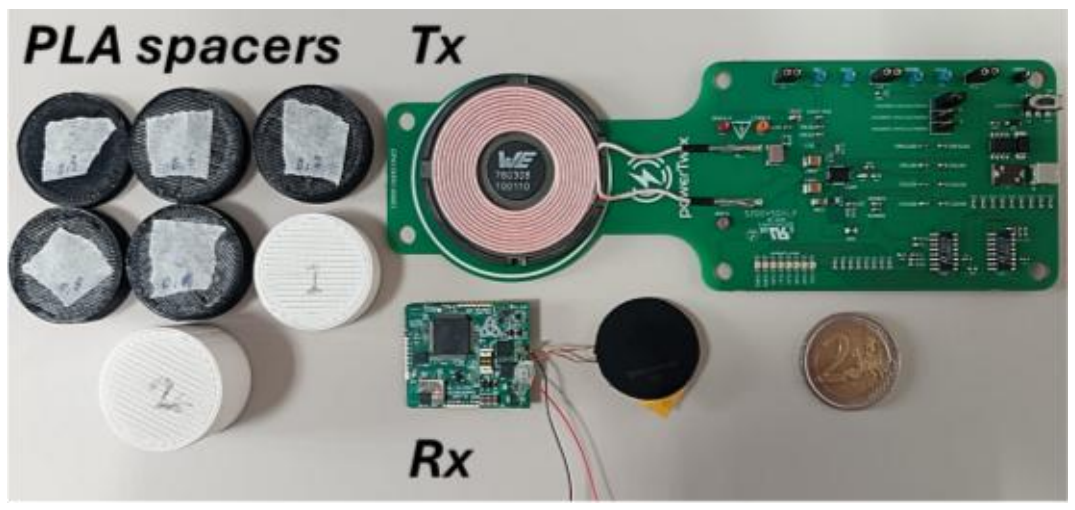

Fig. 6. Photographs of the prototypes and the 3D-printed PLA spacers.

### E. Simulative Safety Check of the WPT

Bodycentric antenna systems must also ensure safety of the radio link operating in the proximity of the human body. The WPT must hence respect the limitations in terms of SAR (Specific Absorption Rate), induced J (electric current density) and induced E (electric field) inside the human tissue [19]. The assessment is simulatively carried out (with CST Microwave Studio 2025, LF module) at the frequency of 127 kHz. The arm is simulated as a cylinder (radius: 40 mm; length: 650 mm) of homogeneous wet skin, which is a conservative approximation for our application (conductivity: 0.082078 S/m; relative permittivity: 12960; values from [20]). The coils' geometries reproduce the real ones, and the Rx coil is simulated as Perfect Electric Conductor, placed directly inside the biological tissue. The Tx coil is simulated with a constant 1 A current on it and the minimum Tx-Rx distance from the project specifications (i.e., 5 mm of minimum implant depth, plus 1 mm of spacing of the Tx from the body accounting for the presence of a case) is considered. Fig. 5(a) shows the CAD employed for the simulation.

Ref. [19] reports the maximum acceptable values for biological safety hereafter used. Figs. 5(b)-(d) depict the simulative results. The SAR averaged on 10 g of homogeneous tissue has a peak value of about 0.326 mW/kg which is a value well below from the basic restriction of 2 W/kg. Similarly, the electrical field peaks at 3.46 V/m in air while the reference limit is of 17.1 V/m RMS (Root Mean Square) inside the biological tissue. The last limiting quantity is the induced current density, which reaches a peak value of 0.206 A/m$^2$ with a corresponding limit of 0.254 A/m$^2$ RMS. Therefore, according to the simulative check, compliance with current regulations is expected for a constant current flowing in the Tx up to 1.74 A, assuming sinusoidal stimulation. However, it is also worth noticing that the CAD model is based on several conservative approximations (homogeneous wet skin, no spacing materials, constant current on the Tx coil), and real-world measurements can prove even higher safety.

## III. Prototypes and Experimentation

The transmitting and receiving custom boards were manufactured by an external service. The transmitting board is a 4-layer, 1.6-mm-height PCB while the receiving one is a 6-layer, 1-mm-height one. For experimentation, the Tx board was powered by a voltage power supply (model CPX400D by Aim TTI; voltage supply fixed at 5 V) providing about 1.2 A, whereas the voltages and currents entering the motor were measured by using an oscilloscope (model HD0610A by Teledyn LeCroy) and a multimeter (model IDM71 by Iso-Tech). Overall, the external size of the Tx unit is 172 mm x 58 mm, compatible with inclusion inside a prosthesis. The size of the Rx board (excluding the motor) is instead 75 mm x 27 mm; the final prototype for implantation will have to shrink the footprint by exploiting the admitted height of 10 mm (see next Sec. IV). 3D-printed, cylindrical spacers in PLA were employed to modulate the distance between the coils during experiments. Fig. 6 shows the prototypes and the spacers. To quantify the efficiency of the prototypes, the value $\eta = P_{load}/P_{source}$ is quantified. Since the current and voltage values are measured at the source and at the load, $\eta$ includes the effects of the electronics and it is named "system efficiency".

TABLE I. Tabular data of the antenna system's performance.

| *Dis. [cm]* | *I [A] (Tx)* | *P [W] (Tx)* | *I [mA] / V[V] (Rx)* | *P [W] (Rx)* | *Sys. Eff. [%]* |
|---|---|---|---|---|---|
| 0.5 | 1.25 | 6.25 | 143 / 19 | 2.73 | 43.5 |
| 0.6 | 1.27 | 6.35 | 124 / 17 | 2.11 | 33.2 |
| 0.7 | 1.30 | 6.50 | 131 / 18 | 2.36 | 36.3 |
| 0.8 | 1.26 | 6.30 | 110 / 16 | 1.76 | 28.0 |
| 0.9 | 1.28 | 6.40 | 108 / 17 | 1.84 | 28.7 |
| 1.0 | 1.19 | 5.95 | 100 / 12 | 1.20 | 20.2 |
| 1.5 | 1.35 | 6.75 | 71 / 9 | 0.64 | 9.5 |
| 2.0 | 1.27 | 6.35 | 20 / 9 | 0.18 | 2.8 |

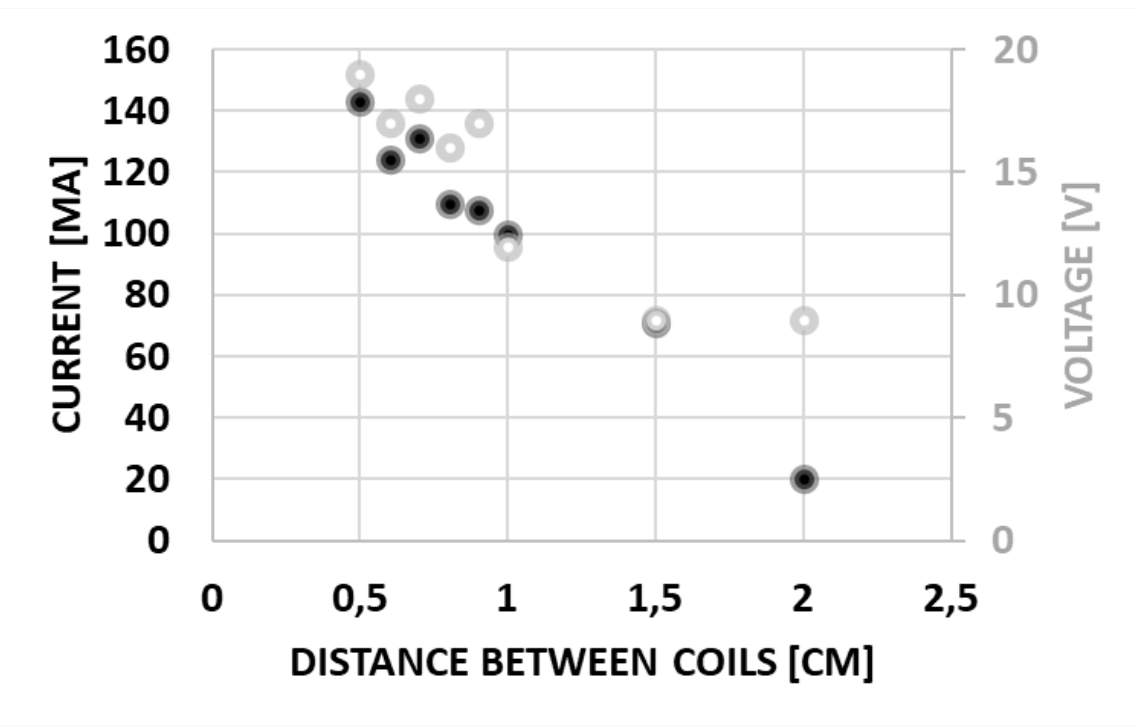

Fig. 7. Current and voltage on the load when the distance between the coil increases.

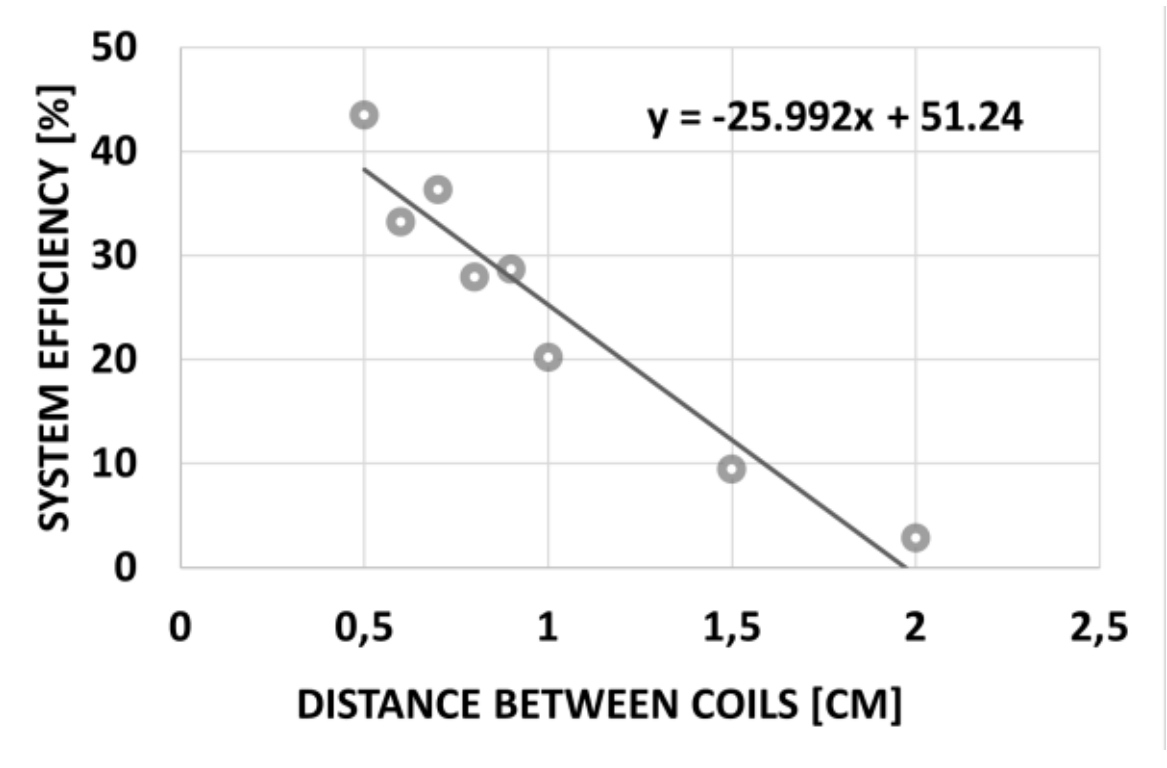

Fig. 8. Relationship between the antenna system's efficiency and the distance between the coils. The equation of the linear regression is also reported.

Table I reports the detailed data including currents, voltages and powers measured. The WPT link satisfies the project requirements all envisioned operations (5 mA and 7 V at the load) up to the maximum implantation depth considered by the project (2 cm; Fig. 7). It is worth noticing that the autoresonance plays a crucial role as the IC manages to stabilize the voltage on the load at 9 V even when the distance between the coils increases from 1.5 cm to 2 cm and the current on the load decreases from 71 mA to 20 mA. Thanks to the autoresonant IC, $\eta$ decreases in a stable, linear way when the distance between the coils increases (Fig. 8). No coupling is established at the next inter-coil testing distance of 2.5 cm, which is beyond the specifics of our bionic interface.

## IV. Discussion and Conclusion

In this contribution, an inductive, LF WPT with autoresonance for multimodal bionic interfaces has been designed and tested, including its compliance with normative regulations. The requirements for the bionic interface were detailed. After the design and simulative checks, preliminary PCB prototypes were manufactured and utilized for testing the antenna system. The WPT tuned at 127 kHz, overall, complies with all the relevant normative and manages to provide up to 20 mA and 9 V when the implantation depth is 2 cm, satisfying the necessary performances for the wireless link.

Based on the preliminary experimentation, the effectiveness of the LF autoresonant WPT for network-connected prosthetics seems suitable. Such cyber-physical prostheses are affected by significant inter- and intra-patient variability, and the automatic adjustments can adequately stabilize the delivered power with variable motion patterns and loads. Such a capability is expected to be vital when integrating the bionic interface, overcoming current antenna systems for implanted WPT [2, 9, 10, 21]. The selected frequency of 127 kHz is also an interesting trade-off between the electromagnetic, electronic, and regulatory (for the European Union) requirements, and it should be considered when designing similar medical systems, in the authors' opinion.

Future work will proceed towards the completion of the bionic interface. On the hardware side, the Rx board respecting the given external footprint must be manufactured, by accepting a reduction of the I and V on the load that, currently, satisfy the requirements by a large margin. The software and firmware governing the electronic components are being developed to provide the needed waveforms and controls in real-time. The safety of the WPT must also be certified through direct measurements. Finally, the inclusion of the Bluetooth communication link is planned to integrate the prosthesis into the patient's WBAN so as to deploy a fully functional prototype of sensing cyberprosthesis [22].

## Acknowledgment

The authors thank Renata De Gennaro, Fabrizio Trentini, and Cosima Fiaschi (all affiliated with EXEMPLAR srl., Turin, Italy) for providing the trial of the CST Microwave Studio software used in this paper.